\documentclass[prc,preprint,showpacs,preprintnumbers,amsmath,amssymb]{revtex4}

\usepackage{graphicx}% Include figure files
\usepackage{dcolumn}% Align table columns on decimal point
\usepackage{bm}% bold math

\begin{document}

\title{Isoscaling bearing information on the nuclear caloric curve}

\author{S.R.\ Souza}
\author{R.\ Donangelo}%
\affiliation{Instituto de Fisica, Universidade Federal do Rio de Janeiro\\
Cidade Universit\'aria, CP 68528, 21941-972 Rio de Janeiro, Brazil}

\author{W.G.\ Lynch}
\author{W.P.\ Tan}
\altaffiliation[Present address:]{Department of physics, University of Notre Dame,
Notre Dame, IN 46556}
\author{M.B.\ Tsang}
\affiliation{Department of Physics and Astronomy and National Superconducting Cyclotron
Laboratory, Michigan State University,\\
East Lansing, Michigan 48824}

\date{\today}% It is always \today, today,

\begin{abstract}
We show that the qualitative behavior of the nuclear caloric curve can be inferred from the
energy dependence of the isoscaling parameters.
Since there are strong indications that the latter are not distorted by the secondary decay
of primary hot fragments, in contrast to other observables, this suggests that
valuable information on the nuclear caloric curve may be obtained through the analysis
presented in this work.
\end{abstract}

\maketitle

The determination of the nuclear caloric curve is key to the understanding of the
multifragment emission observed in heavy-ion collisions at intermediate energies
\cite{ccCampi1996,DasGupta2003}.
Indeed, equilibrium statistical calculations predict a continuous rise of the
breakup temperature as a function of the excitation energy of the disassembling system
if the multifragment emission takes place at fixed density
\cite{smmTan2003,DasGupta2003,isoBondorf1998}.
In this case, the pressure at the breakup stage increases monotonously as a function of the
excitation energy.
On the other hand, a wide plateau should be observed \cite{smm2,isoBondorf1998} if the onset of
multifragmentation occurs after a rapid rise of the pressure, which is accompanied by a steady
decrease of the breakup density as the excitation energy increases \cite{smm2}.
Similar conclusions have also been obtained with a soluble thermodynamic model by Das {\it et al.}
\cite{DasGupta2003}.

The experimental observation of the caloric curve has been intensively debated in the last
years \cite{ccNatowitz2002,ccMoretto1996,ccCampi1996,ccPeter1998,reviewDasGupta2001} since
different measurements have
led to distinctly different conclusions
\cite{ccNatowitz2002,ccRuangma2002,ccKwiatkowski1998,ccgsi1995,ccHauger1997,ccMa1997,cc1998,ccmsu1998}.
One problem is due to the great difficulties in measuring the nuclear temperature as
this quantity can only be inferred from fragment information measured long after the breakup stage.
It has been shown that side feeding from the deexcitation of the primordial hot fragments
\cite{Xi1999,reviewDasGupta2001,ccPeter1998,tempPeter1998,tempdr1996} may bias empirical
conclusions drawn from the different
methods currently employed \cite{tempstates1984,Albergo1985,temp3h3he2001,tempPeter1998} in the
determination of the breakup temperature.

In this context, the isotopic scaling recently observed in nuclear reactions, in a broad range of
bombarding energies
\cite{isoXu2000,isoTsang2001_1,isoTsang2001_2,isoBotvina2002,isoShetty2003,isoSouliotis2003},
is expected to be
rather insensitive to effects associated with the secondary decay of the primary
fragments \cite{isoTsang2001_2,mass2003}.
More precisely, the ratio $R_{21}$ between the multiplicity $Y_i(N,Z)$ of a fragment whose
proton and neutron numbers are, respectively, $Z$ and $N$, measured in two reactions
(1 and 2) with different isospin, follows the relation:
\begin{equation}
R_{21}=\frac{Y_2(N,Z)}{Y_1(N,Z)}=C\exp\left(\alpha N + \beta Z\right)\;,
\label{eq:r21}
\end{equation}

\noindent
where $\alpha$ and $\beta$ are the isoscaling parameters and
$C$ is a normalization constant.
This scaling property is very robust.
As is shown in refs.\ \cite{isoTsang2001_1,isoShetty2003,isoSouliotis2003}, it has been observed in
deep inelastic reactions, evaporation processes, besides nuclear multifragmentation.

One explanation for this scaling behavior may be found in the grand-canonical ensemble, in which the
multiplicity $Y(N,Z)$ is given by:
\begin{equation}
Y(N,Z)=\zeta_{AZ}(T,V_f)\exp\left[\frac{\mu_p Z +\mu_n N}{T}\right]\;,
\label{eq:ygc}
\end{equation}

\noindent
where $T$ is the breakup temperature, $\mu_n$ and $\mu_p$ respectively stand for the neutron and
proton chemical potentials and
\begin{equation}
\zeta_{AZ}=g_{AZ}\frac{V_f}{\lambda_T^3}A^{3/2}\exp\left[\frac{B_{AZ} - f_{AZ}(T)}{T}\right]\;.
\label{eq:fygc}
\end{equation}

\noindent
In the above expression, $A$ is the mass number, $B_{AZ}$ and $g_{AZ}$ are, respectively,
the fragment's binding energy and spin degeneracy factor.
The excitation energy of the fragment is taken into account by the internal free energy $f_{AZ}(T)$.
The free volume $V_f$ is a parameter of the calculation and $\lambda_T=\sqrt{2\pi\hbar^2/m_NT}$,
where $m_N$ is the nucleon mass.

If the breakup takes place in the two reactions at the same temperature and density, the ratio
involving $\zeta_{AZ}$ cancels out and one finds that:

\begin{equation}
\alpha=\frac{\mu^{(2)}_n-\mu^{(1)}_n}{T}\;\;{\rm and}
\;\;\beta=\frac{\mu^{(2)}_p-\mu^{(1)}_p}{T}\;,
\label{eq:alphabeta}
\end{equation}

\noindent
where the superscripts label reactions (1) and (2).
Although the observed yields are affected by the decay of the primary fragments, the form
of Eq.\ (\ref{eq:ygc}) is still expected to hold, even though $\zeta_{AZ}$ will be given by
a more complex expression.
Since $\zeta_{AZ}$ might be very similar in the two reactions, as long as the breakup temperatures
are close enough, the isoscaling parameters should be safely obtained from
the final yields \cite{isoTsang2001_2,mass2003} and, therefore, should bear reliable information
on the breakup stage.

Owing to the exponential relationship between the system mass and the chemical potential,
$\mu^{(2)}-\mu^{(1)}$ should be a slowly varying function of the temperature.
This is indeed confirmed by our calculations (see below), in agreement with the findings of ref.\
\cite{isoBotvina2002}.
Therefore, the robustness of the isoscaling parameters may be used to investigate the qualitative
behavior of the caloric curve.

To address this point, we apply the isoscaling analysis to the decay products of reactions in which a
thermally equilibrated source with excitation energy $E^*$ breaks up statistically.
We consider a proton rich source, whose mass and atomic numbers are $A_0=168$ and $Z_0=75$,
which corresponds to reaction (1).
In the second reaction, we use a neutron rich source, with  $A_0=186$ and $Z_0=75$.
The improved version \cite{smmTan2003} of the statistical multifragmentation model (ISMM) developed
in refs.\ \cite{smm1,smm2,smm3}, and described below, is used in the following calculations to
simulate the decay of the excited source.
Either implementation of this model is useful in the present study since, under different assumptions
for the breakup density, it predicts qualitatively different caloric curves \cite{isoBondorf1998}.
Thus its predictions can provide the input to investigate the sensitivity of the isoscaling
parameters on the qualitative shape of the caloric curve.

In the ISMM model, partitions strictly consistent with the constraints:
\begin{equation}
A_0=\sum_{AZ}N_{AZ}A\;,\;\;\;Z_0=\sum_{AZ}N_{AZ}Z\;,
\label{eq:mcconst}
\end{equation}

\noindent
and
\begin{equation}
E_0^{\rm g.s.}+E^*=\frac{3}{5}\frac{Z_0^2e^2}{R_0}+\sum_{AZ}N_{AZ}E_{AZ}(T,V)\;,
\label{eq:econst}
\end{equation}

\noindent
are imposed.
In the above equations, $N_{AZ}$ denotes the multiplicity, in each generated partition, of fragments
whose mass and atomic numbers are $A$ and $Z$, $E_0^{\rm g.s.}$ is the ground state energy of the
source, $e$ represents the elementary charge, and $R_0$ is the radius of a sphere with a volume $V$,
corresponding to the breakup volume.
The energy $E_{AZ}(T,V)$ contains contributions from the fragment's binding energy, excitation energy,
translational motion, besides the remaining Coulomb terms which, through the Wigner-Seitz
approximation, provide the corrections to account for the Coulomb repulsion between the fragments
\cite{smm1}.
A Monte Carlo sample of the possible fragmentation modes is carried out following \cite{smm3}.
The breakup temperature is determined, for each partition, by solving Eq.\ (\ref{eq:econst}).

The main differences from the ISMM \cite{smmTan2003} and the original SMM \cite{smm1,smm2,smm3} are
in the use of internal free energies built from empirical data on discrete states wherever
available and of experimental binding energies all over the mass table \cite{awmtable1995}.
Careful extrapolations are carried out to mass regions where the information is not available in
either case \cite{smmTan2003,mass2003}.
Both quantities influence directly the determination of $T$ in each fragmentation mode $f$,
whereas the free energies also play an important role in the evaluation of the entropy
$S_f$, which enters in the calculation of any physical observable $O_{AZ}$:
\begin{equation}
\langle O_{AZ}\rangle = \frac{\sum_fO_{AZ} \exp\left[\sum_{\{AZ\}_f}N_{AZ}S_{AZ}\right]}
                        {\sum_f\exp\left[\sum_{\{AZ\}_f}N_{AZ}S_{AZ}\right]}
%\langle O_{AZ}\rangle = \frac{\sum_f\exp\left[\sum_{\left{AZ\right}}N_{AZ}S_{AZ}\right]O_{AZ}}
%                             {\sum_f\exp\left[\sum_{\left{AZ\right}}N_{AZ}S_{AZ}\right]}\;.
\label{eq:observable}
\end{equation}

\noindent
Due to the constraints imposed on each partition, a physical observable $O_{AZ}$ fluctuates from
one fragmentation mode to the other.

\begin{figure}
\includegraphics[angle=0,totalheight=7.0cm]{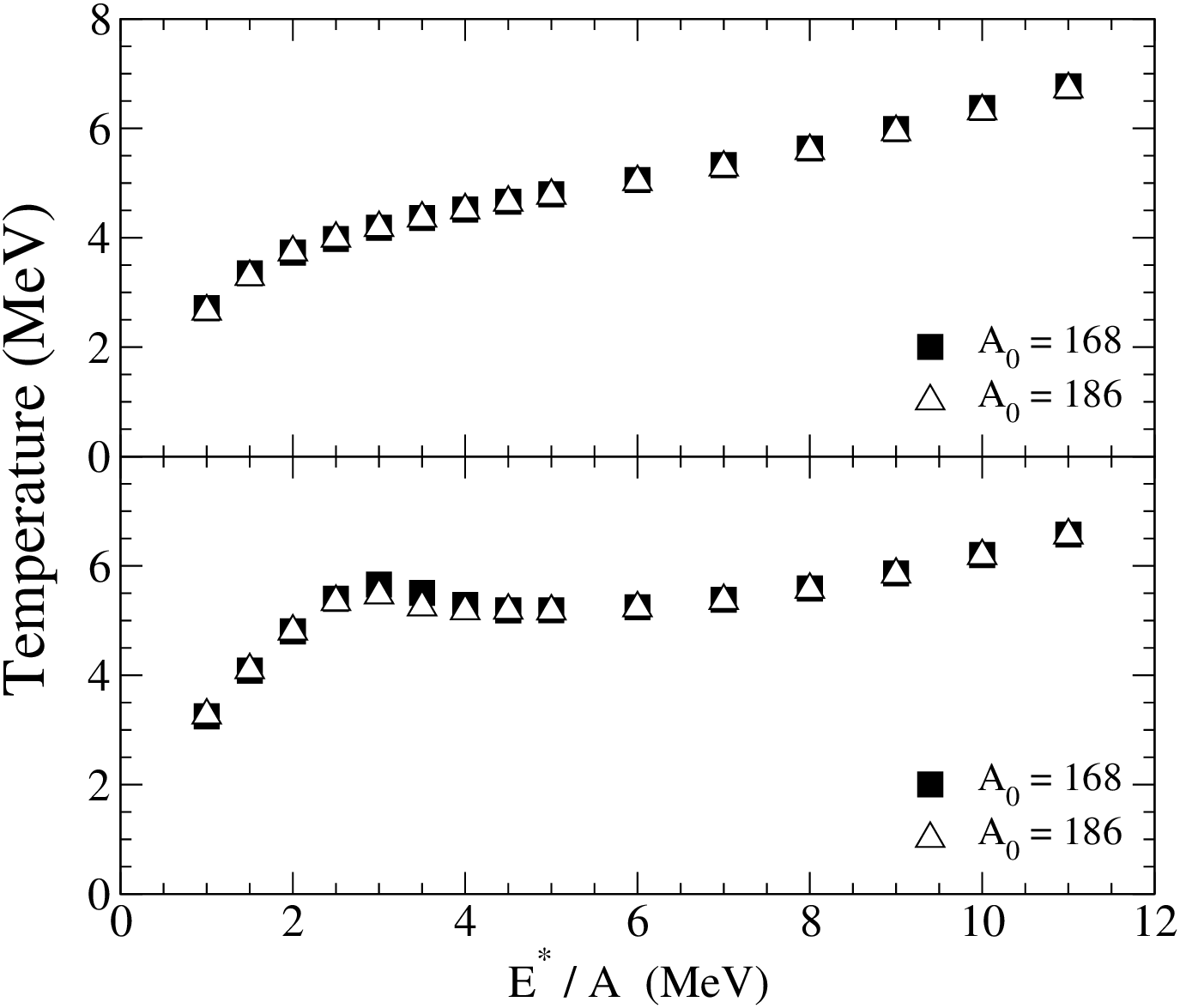}
\caption{\label{fig:cc}Caloric curve predicted by the ISMM for the neutron (triangles)
and proton (squares) rich sources. In the upper panel it is assumed that the breakup occurs
at a fixed density whereas the breakup volume is allowed to change in the results displayed
in the bottom panel. For details see text.}
\end{figure}

The average breakup temperature may be obtained through the above expression and it is shown
in Fig.\ \ref{fig:cc} as a function of the excitation energy for the two sources we consider.
In order to suppress statistical fluctuations, a billion events have been generated 
for each excitation energy.
A monotonous increase of the temperature is observed in the upper panel, in which case the
breakup volume of the system is kept constant and is 6 times larger than its value at normal density.
One also observes that the average temperatures are very similar for both sources, which justifies
the assumption that the temperature dependent terms in $\zeta_{AZ}$ cancel out in
Eq.\ (\ref{eq:r21}).
As stated before, and reported in ref.\ \cite{isoBondorf1998}, the situation is qualitatively
different as one allows the breakup volume to be
multiplicity dependent and a plateau is observed approximately
between 3.5 and 7 MeV/A.
The breakup temperatures in both systems are quite close, except at the region around the
onset of multifragmentation, $E^*/A \approx 3.0$ MeV, where small differences may be noticed.

In contrast to an earlier study also using the SMM \cite{ccBotvina2002}, in which a larger isospin
dependence in the plateau region was predicted, such effect is rather reduced in
our calculations.
To understand this aspect, we rewrite Eq. (\ref{eq:econst}) in the following way:

\begin{equation}
E^*-\left[B_{A_0Z_0}-\sum_{AZ}N_{AZ}B_{AZ}\right]
-\left[\frac{3}{5}\frac{Z_0^2e^2}{R_0}-\sum_{AZ}N_{AZ}\frac{3}{5}\frac{Z^2e^2}{R_{cell}}\right]
=\frac{3}{2}T(M-1)+\sum_{A}N_AE^*_A(T)\;.
\label{eq:temp}
\end{equation}

\noindent
where $N_A$ denotes the multiplicity of fragments with mass $A$ in
the fragmentation mode considered and the total fragment multiplicity is represented by $M$.
The quantity $R_{cell}$ stands for the radius of the cell in which the fragment
is embedded as one subdivides the system in order to apply the Wigner-Seitz approximation \cite{smm1}.

The first term on the right hand side of Eq.\ (\ref{eq:temp}) corresponds to the fragments'
kinetic energy (in the center of mass reference frame of the total system) and $E^*_A(T)$ represents
the internal excitation energy of a fragment with mass $A$ at temperature $T$.
We suppressed any $Z$ dependence in $E^*_A(T)$ since it has been introduced in SMM only
in ref.\ \cite{smmTan2003}.
It entered in previous calculations only in the case of $A=4$ nuclei because all light
fragments ($A<5$) were assumed to have no internal degrees of freedom, except for the alpha
particles.
Therefore, all the dependence on isospin is confined to the left hand side of this
equation.

An increase of $Z_0/A_0$ enhances the Coulomb term between brackets in Eq.\ (\ref{eq:temp}) and, as a
consequence, tends to lower the temperature, which appears only on the right hand side of the
above expression.
This is the effect observed in ref.\ \cite{ccBotvina2002}.

Nevertheless, the binding energies play a very important role in the balance
of the equation and, consequently, on the temperature extracted from it.
It is shown in ref.\ \cite{mass2003} that {\it total} binding energies calculated through simple
liquid drop mass formulae, such as that used in ref.\ \cite{ccBotvina2002}, deviate appreciably
from empirical values.
The differences reported in ref.\ \cite{mass2003} can be as large as 40 MeV for heavy nuclei
and are, on the average, around 10 MeV for light fragments.
Therefore, $B_{A_0Z_0}-\sum_{AZ}N_{AZ}B_{AZ}$ calculated in \cite{ccBotvina2002} has systematic
errors which further increase the isospin dependence of the temperature.
Our results, which are obtained with empirical binding energies and careful extrapolations
to mass regions where this information is not avaliable, as described in \cite{mass2003,smmTan2003}, 
exhibit a much weaker isospin dependence.

\begin{figure}
\includegraphics[angle=0,totalheight=7.0cm]{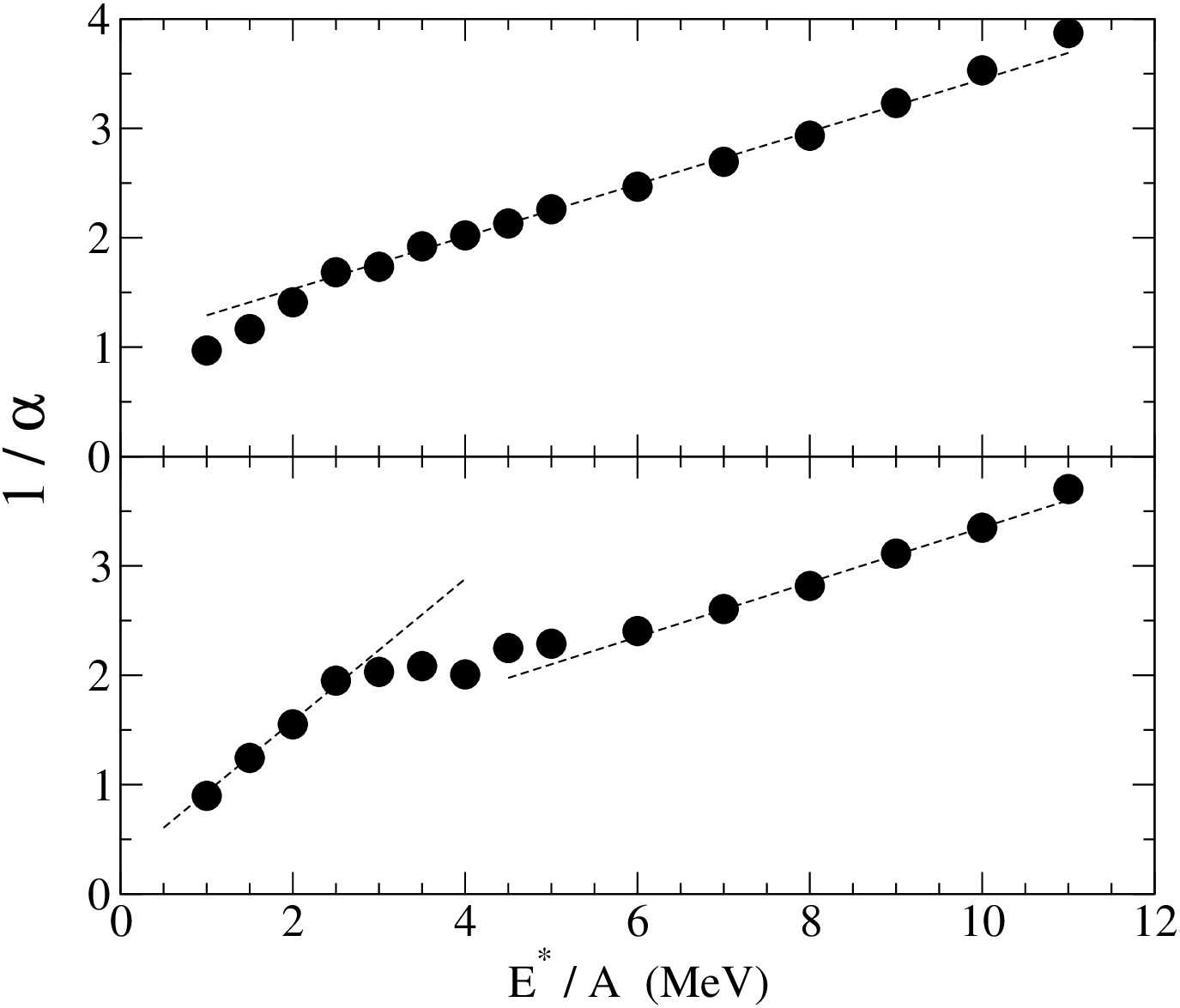}
\caption{\label{fig:ainv}Reciprocal of the isoscaling parameter $\alpha$ as a function of
the excitation energy.
Constant breakup volume is assumed in the upper panel, while in the lower one the
results were obtained with variable breakup volume.
The lines illustrate the slopes in each region. For details see text.}
\end{figure}

We now investigate the extent to which the isoscaling parameters carry informaton on the
caloric curve and show, in Fig. \ref{fig:ainv}, $1/\alpha$ as a function of the
excitation energy.
The parameters are obtained by fitting Eq.\ (\ref{eq:r21}) to the primary yields
predicted by the microcanonical ISMM calculations, with the same isotopes, $Z = 1$, 2,$\dots$, 8,
used in ref.\ \cite{isoTsang2001_2}.
We have checked that, in agreement with \cite{isoTsang2001_2,mass2003}, the changes in
$1/\alpha$ due to the secondary decay \cite{smmTan2003} of the primordial fragments are of the order
of 5\% at $E^*/A = 3$, 6, and 9 MeV and, therefore, do not change our conclusions.
As in the previous plot, the results shown in the upper panel correspond to a fixed breakup
density whereas it is allowed to change in the lower part of the picture.
The results reveal distinct qualitative behaviors in each case.
More specifically, the reciprocal of $\alpha$ follows approximately a straight line for almost
the full range of excitation energies considered if the breakup density is kept fixed.
A clear change of slope, before and after the plateau region, appears if the breakup volume is
multiplicity dependent.
The plateau observed in the caloric curve is not so apparent in the $1/\alpha$ plot because,
although $\mu^{(2)}-\mu^{(1)}$ varies slowly compared to $T$, it somewhat distorts the curve.
The relevant point here is that one should observe different qualitative behaviors in the
$1/\alpha$ plot according to the characteristics of the caloric curve.
Similar conclusions are obtained with the $\beta$ parameter, but we concentrate on $\alpha$ because,
as shown in refs.\ \cite{isoTsang2001_2,mass2003}, $\beta$ might change more than $\alpha$ after
secondary decay of the primary fragments, probably due to Coulomb effects.
The distinction between the two scenarios allows one to ascertain the existence of the
plateau in the caloric curve.
It is worth mentioning that the experimental results presented in \cite{isoBotvina2002} seem to
favour the monotonous increase of the breakup temperature.
However, since in this reference $1/\alpha$ is plotted as a function of the bombarding
energy, further analysis is needed to draw more precise conclusions.

In order to investigate the dependence of the results on the statistical ensemble employed,
we also calculate the primary yields, for the two sources considered here, in the
framework of the grand-canonical approach.
To prevent artificial deviations, we use the same ingredients of the micro-canonical case,
such as binding energies, internal free energies, and spin degeneracy factors.
The free volume and the breakup temperature, which enter into eqs.\ (\ref{eq:ygc}) and
(\ref{eq:fygc}), are obtained from the micro-canonical calculation for each excitation energy,
instead of considering them as free parameters as is usually assumed.
Given these two quantities, the chemical potentials are calculated by imposing that:

\begin{equation}
A_0=\sum_{NZ}Y(N,Z)A\;\;\; {\rm and}\;\;\; Z_0=\sum_{NZ}Y(N,Z)Z\;,
\label{eq:constgc}
\end{equation}

\noindent
where $Y(N,Z)$ is computed through eqs.\ (\ref{eq:ygc}) and (\ref{eq:fygc}).
The chemical potentials are found by minimizing the difference between the left and the right
hand sides of the above equations.
We impose that the constraints are fulfilled with a precision better than 3 digits.
Once the chemical potentials are obtained, the primary yields $Y(N,Z)$ may be evaluated through
eqs.\ (\ref{eq:ygc}) and (\ref{eq:fygc}).

\begin{figure}
\includegraphics[angle=0,totalheight=7.0cm]{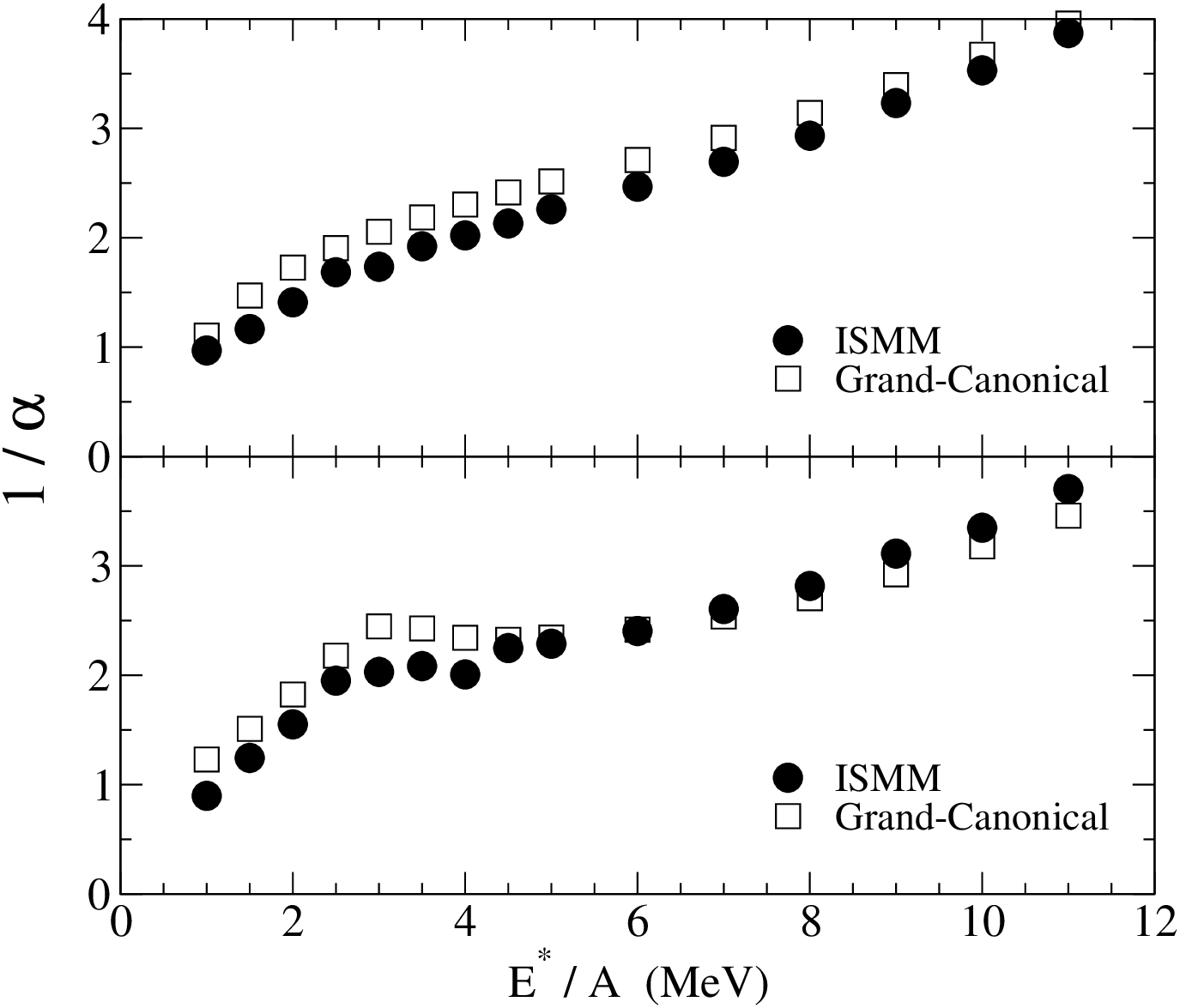}
\caption{\label{fig:amcgc}Same as fig.\ \ref{fig:ainv}.
The circles represent the predictions of the micro-canonical approach, whereas the
results obtained with the grand-canonical ensemble are depicted by the squares.
For details see text.}
\end{figure}

The predictions of the two ensembles are compared in Fig.\ \ref{fig:amcgc}, where the
results of the grand-canonical calculations are depicted by the open squares whereas
the full circles represent those obtained with the micro-canonical ensemble.
The isoscaling parameters are obtained by fitting Eq.\ (\ref{eq:r21}) using the grand-canonical
yields, as is done in the micro-canonical calculation.
It may be noticed that, although the absolute values differ from one statistical approach
to the other, both exhibit the same qualitative behavior.
We have checked that $\alpha$ and $\beta$ obtained through Eq.\ (\ref{eq:alphabeta}) agree fairly
well with the grand-canonical values displayed in this picture.
Small deviations are observed only in the regions in which the difference between the breakup
temperatures of the sources are non-negligible.
Besides showing the consistency of the isoscaling assumptions, this confirms the expectation
that small differences between the average temperatures of the two sources should not appreciably
affect the scaling properties.

The difference between the predictions of the two statistical approaches may be attributed to
the strong mass, charge, and energy constraints imposed in the micro-canonical calculations in
each partition.
These restrictions lead to a fairly broad temperature distribution \cite{tempfluct2000}, corresponding
to different fragmentation modes.
In contrast, the temperature is kept constant in the grand-canonical ensemble whereas the
charge, mass and energy are fixed only on the average and not event by event.
However, the qualitative agreement between the two calculations is a very positive aspect since
it is extremely difficult to select events whose decaying sources strictly obey these
constraints in experiments.
Therefore our analysis seem to indicate that, at least, the qualitative behavior of the caloric
curve may be studied experimentally.

In conclusion, we suggest that the existence of the plateau in the nuclear caloric curve may
be better investigated experimentally through the isoscaling analysis.
Clear deviations from the linear behavior are expected to be found in the $1/\alpha$ vs $E^*/A$
curve if the plateau exists.
Conversely, if this plateau does not exist, the $1/\alpha$ vs $E^*/A$ curve should follow a straight
line over a wide excitation energy domain.
Our results indicate that the determination of the qualitative shape of the caloric curve
can be done more reliably using the isoscaling analysis than through measurements of the
temperature from the multiplicities of the detected fragments.

\begin{acknowledgments}
We would like to acknowledge CNPq and FUJB for partial financial support.
This work was supported in part by the National Science Foundation under Grant
No.\ PHY-01-10253 and INT-9908727 and by the CNPq-NSF agreement.
\end{acknowledgments}

\bibliography{cc_iso}

\end{document}